\documentclass[%
reprint,
superscriptaddress,
amsmath,amssymb,
prl,
prstper,
floatfix,
]{revtex4-1}

\usepackage{upgreek}
\usepackage{soul}
\usepackage{color}
\usepackage{graphicx}
\usepackage{dcolumn}
\usepackage{bm}

\begin{document}

\title{Formation of Bound States in the Continuum in Hybrid Plasmonic-Photonic Systems}
\author{Shaimaa I. Azzam}
\email{sazzam@purdue.edu}
\affiliation{School of Electrical and Computer Engineering and Birck Nanotechnology Center, Purdue University, West Lafayette, IN 47907, USA}

\author{Vladimir M. Shalaev}
\email{shalaev@purdue.edu}
\affiliation{School of Electrical and Computer Engineering and Birck Nanotechnology Center, Purdue University, West Lafayette, IN 47907, USA}

\author{Alexandra Boltasseva}
\email{aeb@purdue.edu}
\affiliation{School of Electrical and Computer Engineering and Birck Nanotechnology Center, Purdue University, West Lafayette, IN 47907, USA}

\author{Alexander V. Kildishev}
\email{kildishev@purdue.edu}
\affiliation{School of Electrical and Computer Engineering and Birck Nanotechnology Center, Purdue University, West Lafayette, IN 47907, USA}

\begin{abstract}
A bound state in the continuum (BIC) is a localized state of an open structure with access to radiation channels, yet it remains highly confined with, in theory, infinite lifetime and quality factor. There have been many realizations of such exceptional states in dielectric systems without loss. However, realizing BICs in lossy systems such as those in plasmonics remains a challenge. In this Letter, we explore the possibility of realizing BICs in a hybrid plasmonic-photonic structure consisting of a plasmonic grating coupled to a dielectric optical waveguide with diverging  radiative quality factors. The plasmonic-photonic system supports two distinct groups of BICs: symmetry protected BICs at the $\Gamma$-point and off-$\Gamma$ Friedrich-Wintgen BICs. The photonic waveguide modes are strongly coupled to the gap plasmons in the grating leading to an avoided crossing behavior with a high value of Rabi splitting of 150 meV. 
Moreover, we show that the strong coupling significantly alters the band diagram of the hybrid system revealing opportunities for supporting stopped light at an off-$\Gamma$ wide angular span.

\end{abstract}

\pacs{Valid PACS appear here}
\maketitle

In 1929,  von Neumann and Wigner predicted the existence of localized eigenstates of the single particle Schr{\"o}dinger equation embedded in the continuum of eigenvalue state solutions, now known as embedded eigenstates or bound states in the continuum (BICs) \cite{von1929some}. 
Later, there have been many explanations for the formation mechanisms leading to the different types of BICs including symmetry-incompatibility \cite{lee2012observation, plotnik2011experimental, moiseyev2009suppression}, and destructive interference of resonances \cite{hsu2016bound, noda2014analytical, liu2009resonance, gao2016formation, Sadreev2008bound}.
According to Friedrich and Wintgen, when two resonances pass each other as a function of a continuous parameter, the two channels interfere resulting in an avoided crossing of their resonances. Typically, at a given value of the continuous parameter, one of the channels vanishes entirely and hence becoming a BIC with an infinite quality (Q) factor \cite{friedrich1985interfering}. With any perturbation in the ideal system, the BIC would collapse to a Fano resonance with a finite lifetime -- a regime known as quasi-BIC \cite{sadrieva2017transition, hsu2016bound}.
Generally, practical realizations are limited to the quasi-BIC regime as, theoretically, true BICs can only be achieved in systems with at least one dimension extending to infinity \cite{hsu2016bound}.

Another imperative regime is so-called ``near-BIC" at which very high Q factors are still attainable in the vicinity of the BIC.
These high-Q resonances can be explained using the Theory of Resonance Reactions by Fonda \cite{fonda1961resonance, fonda1963bound}; a bound state can exist in the continuum at an energy where some channels are open even when the open and closed channels are strongly coupled.  If the actual Hamiltonian differs inconsiderably from the one which would produce such  bound state, a sharp resonance arises in all scattering and reaction cross sections.
Central advantages of the near-BIC regime are that, unlike regular guided modes below the light-line, the modes in this regime can be excited by free propagating plane waves. Moreover, near-BIC resonances can be obtained over an interval of the continuous parameter values, in contrast to BICs which are generally achieved at a single value of the parameter. This provides stability to fabrication tolerances and other imperfections \cite{sadrieva2017transition}. 

BICs in photonic systems were realized for the first time in 2008 \cite{marinica2008bound} and since then there has been growing interest in their implementations as well as practical applications \cite{gomis2017anisotropy, spinorbit2017, enz2017, bulgakov2017topological, xiao2017topological, disorder2018, alu2018experimental, giantnonlinear2018,monticone2018trapping,lee2012observation,segev2017topologically, Arseniy2018}. There have been numerous realizations of BICs in dielectric systems [\cite{kivshar2017high, bulgakov2017topological, hsu2013observation, kante2017lasing,lee2012observation, alu2018embedded}. However, metallic structures that support BICs have remained limited due to their intrinsic losses \cite{alu2018embedded}.
Proposals for realizing BICs in systems with metallic components have been restricted to either working at the limit of vanishing metal losses \cite{monticone2014embedded} or compensating the losses with gain from an active medium \cite{alu2018embedded}.
Very recently, realization of BICs in hybrid  transition metal dichalcogenides and a dielectric photonic crystal slab has been studied \cite{koshelev2018strong}.
Extensive efforts have been made to realize high-Q plasmonic systems \cite{whispering2009, thackray2015super, oulton2009plasmon, oulton2009plasmon, jelena2007}. Plasmonic cavities with relatively high Q factors have been achieved based on defect modes \cite{jelena2007}, collective plasmon resonances \cite{thackray2015super}, metallic trench Fabry-Perot resonator \cite{zhu2017surface}, high aspect ratio metallic mirrors \cite{sorger2009plasmonic}, or plasmonic nanocavities \cite{noris2017}. However, the demonstrated Q factors of the plasmonic cavities have thus far been limited to a few hundred due to both the radiation and metallic losses.

\begin{figure}[t] \centering
 	\includegraphics[scale= 1.00]{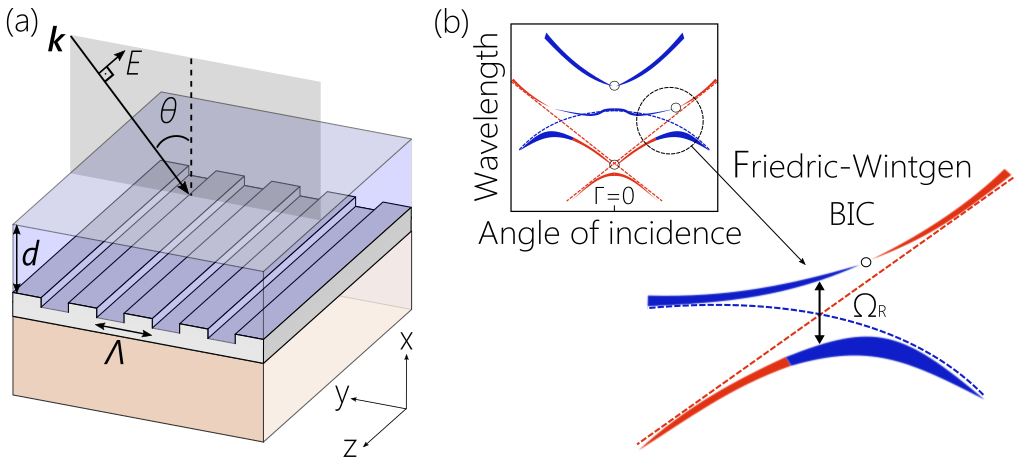}
	 \caption{BICs due to the coupling of plasmon polaritons with photonic modes. (a) A hybrid plasmonic-photonic structure with a silver relief grating (pitch $\Lambda$ = 400 nm) coupled to a $\text{SiO}_2$ slab (thickness $d$ = 500 nm). The periodic ridges of the grating are 30-nm high and 100-nm wide. The back silver film is 100-nm thick. (b) A sketch of the band structures of the hybrid system with coupled plasmonic (blue) and photonic (red) resonances. Dotted lines indicate the unperturbed resonances and solid ones are the coupled resonances. Small circles show the BIC points, and the large dotted circle highlights the strong coupling with Rabi splitting ($\Omega_R$) and showing the off-$\Gamma$ Friedrich-Wintgen BIC.
	 	 }
\label{fig:fig1} 
\end{figure}

\begin{figure*}[t] \centering
    \includegraphics[scale=0.99]{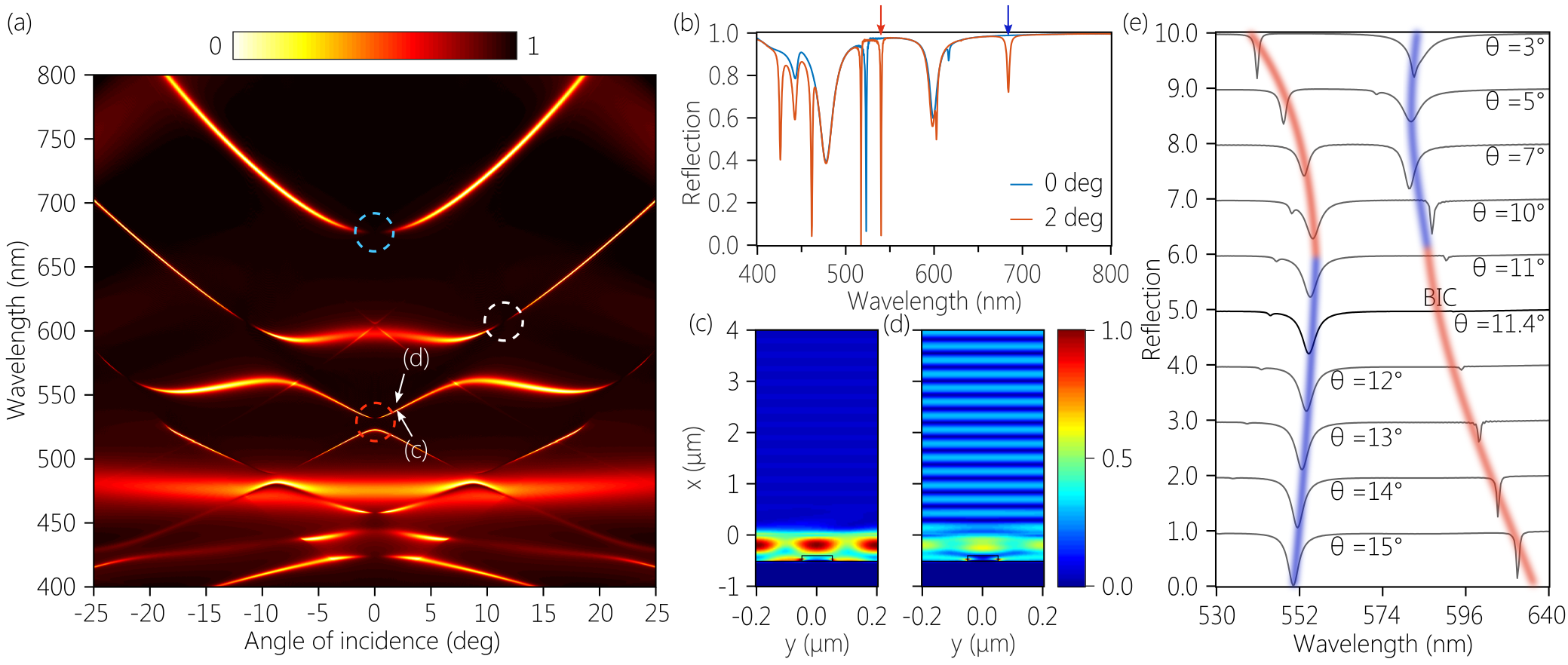}
    \caption{BIC formation in the hybrid plasmonic-photonic system. (a) Reflection spectrum as a function of the angle of incidence (band diagram) of the coupled system. BICs are highlighted by dotted circles showing two symmetry protected BICs due to plasmonic (blue) and photonic (red) modes and Friedrich-Wintgen BIC due to the interference of plasmonic and photonic resonances (white). (b) Cross-sections of the reflection coefficient at $0 ^\circ$ and $2 ^\circ$ incidence angles show the appearance of the collapsed symmetry protected BICs to sharp Fano resonances. Blue and red arrows point to the plasmonic and the photonic collapsed symmetry-protected BICs. Normalized electric fields at $2 ^\circ$ at (c) 540.2 nm (at the resonance) and (d) 541.2 nm (1 nm away from the resonance) showing the filed localization. (e) Avoided crossing and linewidth vanishing due to the hybridization of plasmonic and photonic modes at different angles of incidences. 
   }
\label{fig:fig2} 
\end{figure*}
In this work, the radiative losses in plasmonic structures have been entirely eliminated by the judicious design of a structure that strongly couples surface plasmon polaritons to photonic modes to form BICs. The two interacting channels in the system are: the photonic modes of a dielectric slab waveguide and the plasmonic modes supported by a one-dimensional (1D) periodic silver relief grating, Fig.~\ref{fig:fig1}(a). When the two channels strongly couple, two distinct features can be observed. First, an avoided crossing of the two angular dispersion lines with a reasonably large Rabi splitting -- a signature of the strong coupling between the modes leading to the formation of a polaritonic-optical quasi-particle  \cite{quasiparticle2003, quasiparticle2005}. Second, the vanishing of one of the spectral lines at a specific angle of illumination -- a distinct feature of the BIC formation caused by the destructive interference of the resonances.

A schematic of the resonances’ interactions is depicted in Fig.~\ref{fig:fig1}(b). The unperturbed resonances of the plasmonic (blue) and photonic (red) modes are shown as dotted lines while the coupled modes of the hybrid system are depicted as solid lines. The angular dispersion of surface plasmons excited on the surface-relief grating is a result of a hybridization of the classical surface plasmon-polariton with a plasmonic grating. Subsequently, a band-gap opens up at a wavelength determined by the grating periodicity \cite{barnes1996}:

\begin{equation}
\kappa_{sp} = \frac{2\pi n_\text{eff}}{\lambda_0} = \frac{2\pi}{\lambda_0} \text{sin}\,\theta \pm \frac{2\pi m}{\Lambda},\;  (m = 0, 1,\dots)
\end{equation}

\noindent
 where $\kappa_{sp}$ is the momentum of the surface plasmon, $\lambda_0$ is the free space wavelength, $n_\text{eff}$ is the effective refractive index of the plasmonic mode, $\theta$ is the angle of incidence, $\Lambda$ is the grating pitch, and $m$ is the grating diffraction order.
 
 Likewise, the hybridization of the dielectric slab waveguide with the grating results in the excitation of the photonic modes with a narrow spectral width  \cite{fan2010fundamental}. 
 Overall, the hybrid system supports two distinct types of BICs: symmetry-protected and Friedrich-Wintgen BICs. The former appears at the $\Gamma$-point at normal incidence due to the symmetry incompatibility with the outgoing fields, leading to the decoupling from the free-space propagating waves and hence the localization of the state. We denote the BIC originating from the symmetry of the plasmonic modes as symmetry-protected plasmonic BIC, while that originating from the symmetry of the photonic modes as symmetry-protected photonic BIC. Friedrich-Wintgen BIC appears at an off-$\Gamma$ point due to destructive interference of the resonances at the correct phase matching conditions \cite{friedrich1985interfering}. This type of bound states is also sometimes referred to as an ``accidental'' BIC \cite{hsu2013observation}. 
 In the absence of strong coupling, the dispersion of the photonic and the plasmonic modes will merely cross each other at a particular angular position. However, due to the strong coupling of the modes, an avoided resonances crossing is observed with a Rabi splitting  $\Omega_R$, Fig.~\ref{fig:fig1}(b). This is a signature of the hybridization of the plasmon polaritons and the optical waveguide mode leading to the formation of a new polaritonic-optical quasi-particle \cite{quasiparticle2003, quasiparticle2005} as well as the BIC. An essential characteristic of the Friedrich-Wintgen BIC achieved in the proposed structure is that it originates from the interference of two resonances with different natures, namely a photonic mode and a plasmon polaritons. This provides high flexibility in the engineering of the resonances' dispersion  enabling high control on their spectral positions, angular span, and linewidth.
 
 \begin{figure}[t] \centering
    \includegraphics[scale=1]{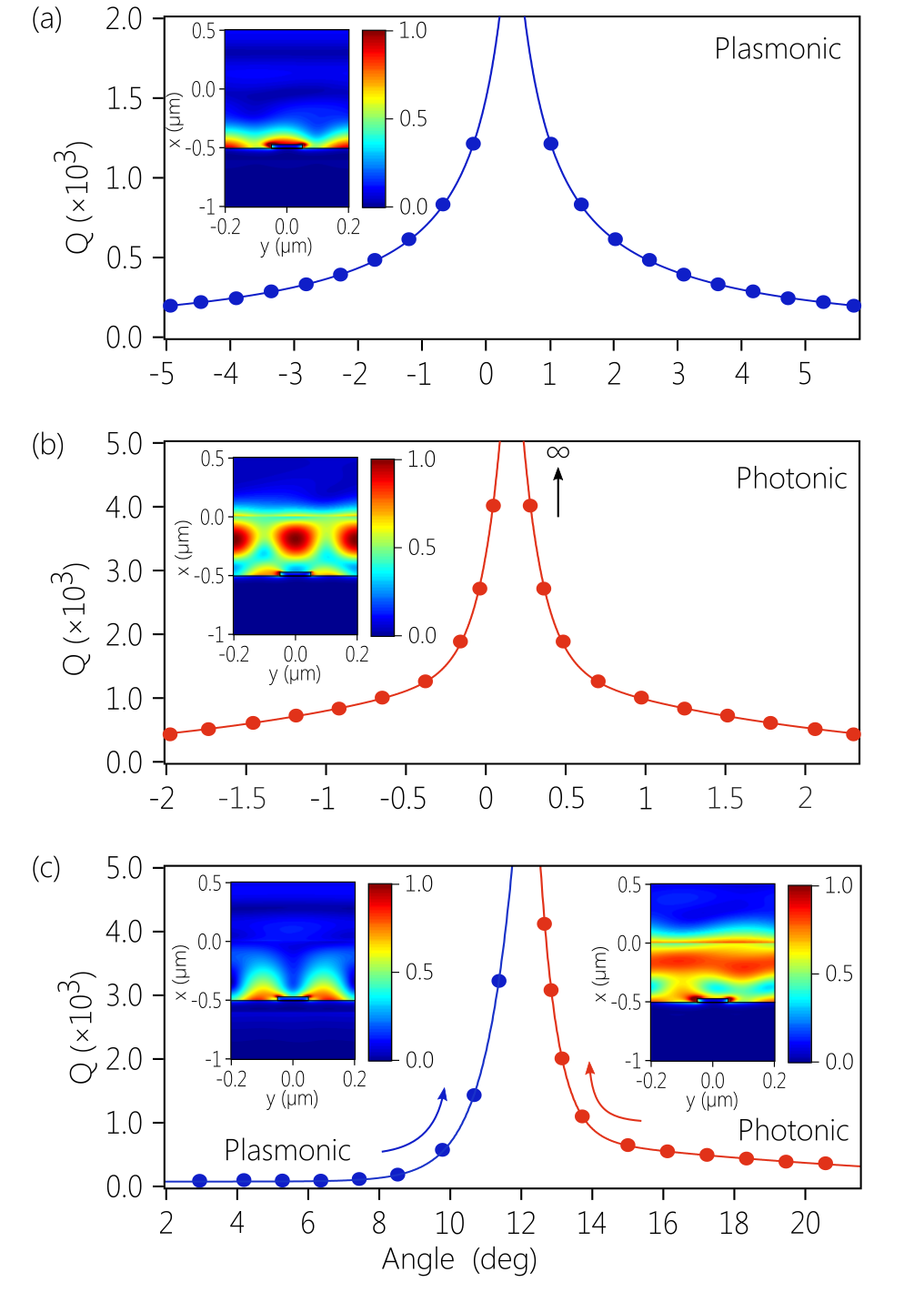}
    \caption{Quality factor vs. angle of incidence for different BIC types. Q factor rapidly increasing towards the $\Gamma$-point for (a) plasmonic and (b) photonic resonances. Insets show the electric fields at (a) 697.1 nm at $5 ^\circ$ and (b) 540.1 nm at $2 ^\circ$. (c) Friedrich-Wintgen BIC appears at $11.4 ^\circ$. Insets show fields of the plasmonic channel at $5 ^\circ$ 596.3 nm (left) and the photonic one $15 ^\circ$ 630.3 nm (right).}
    \label{fig:fig3}     
\end{figure}
 The proposed structure is composed of a silver relief grating (with pitch $\Lambda$ = 400 nm) on top of a silica glass substrate substrate coupled to a silicon dioxide ($\text{SiO}_2$) slab with thickness $d $  = 500 nm. The geometrical parameters are chosen to ensure spectral overlap of the plasmonic resonances and the photonic modes to achieve coupling. 
 The periodic ridges of the grating are 30-nm high and 100-nm wide while the back silver film is 100-nm thick. The dielectric constant of $\text{SiO}_2$ is fixed at 2.25, and the permittivity of the silver is taken from ellipsometric measurements (see Supplemental Material (SM) \cite{SM}). 
 The simulated reflection spectrum as a function of the angle of incidence (band diagram) is shown in Fig.~\ref{fig:fig2}(a).  We use finite difference time domain method for calculating band diagrams and electromagnetic fields; more details can be found in SM \cite{SM}. In this study, BICs are characterized as features of the transmission spectrum at certain angles (wavevectors). However, it is worth mentioning that BICs are the eigensolutions to the source-free Maxwell equations characterized by real eigenvalues above the lightline \cite{bulgakov2014bloch}. Therefore, the angle of incidence is not associated with a source, but rather with the wavevector in the solution.
 The symmetry-protected plasmonic and photonic BICs are shown in Fig.~\ref{fig:fig2}(a) inside blue and red dotted circles, respectively. Both BICs appear at the normal incidence. More notably, the Friedrich-Wintgen BIC, inside a white circle, resulting from the interference and the strong coupling of the photonic and the plasmonic modes, appears at $11.4 ^\circ$ angle of incidence. The calculated reflection coefficients are plotted at angles $0 ^\circ$ and $2 ^\circ$, Fig.~\ref{fig:fig2}(b), demonstrating the noticeable and sudden appearance of sharp Fano resonances when moving away from the symmetry-protected BIC. This is a clear manifestation of the theory of resonances when operating at the near-BIC regime \cite{fonda1961resonance, fonda1963bound}. Also, vertical cross-sections of the electric field at the near-BIC point at 540.2 nm at $2 ^\circ$ is shown in Fig.~\ref{fig:fig2}(c) to verify in-plane light confinement. The localization of the field inside the dielectric layer and minimal radiation is  visible even with available access to the radiation channels. However, even a small deviation (by just 1 nm) from the near-BIC resonance of Fig.~\ref{fig:fig2}(d) would result in significant radiation to the air and nearly one order of magnitude drop in the max field enhancement (not shown). 
The ability to confine and enhance light at the near-BIC is very crucial for enhancing light-matter interaction. However, more importantly, light confined at this regime, unlike regular guided modes below the light-line, can be excited by the free propagating plane wave. The reflection coefficients at selected angles of incidence are scaled and plotted in Fig.~\ref{fig:fig2}(e) illustrating an avoided crossing behavior with Rabi splitting of 150 meV. The natures of the modes are exchanged due to the strong coupling, and at an angle of $11.4 ^\circ$, the spectral width of one resonance entirely vanishes. 
The band diagram of the silver grating covered with an extended $\text{SiO}_2$ layer is given in Fig. S3. It shows the dispersion of only plasmonic modes due to the absence of mode hybridization and strong coupling. Also, the nature of the plasmonic component is explored by calculating the band diagram of a high-index dielectric grating on top of a silver back-reflector, Fig. S5. The band diagram still shows narrow resonances due to the excitation of the photonic modes in the top waveguiding layer due to the additional momentum provided by the grating. However, we observe resonance crossing and no strong coupling in the system which suggests that the origin of surface plasmon modes involved in the coupling is not the propagating surface plasmon but the gap plasmon \cite{SM}. 

\begin{figure}[t] \centering
    \includegraphics[scale=1]{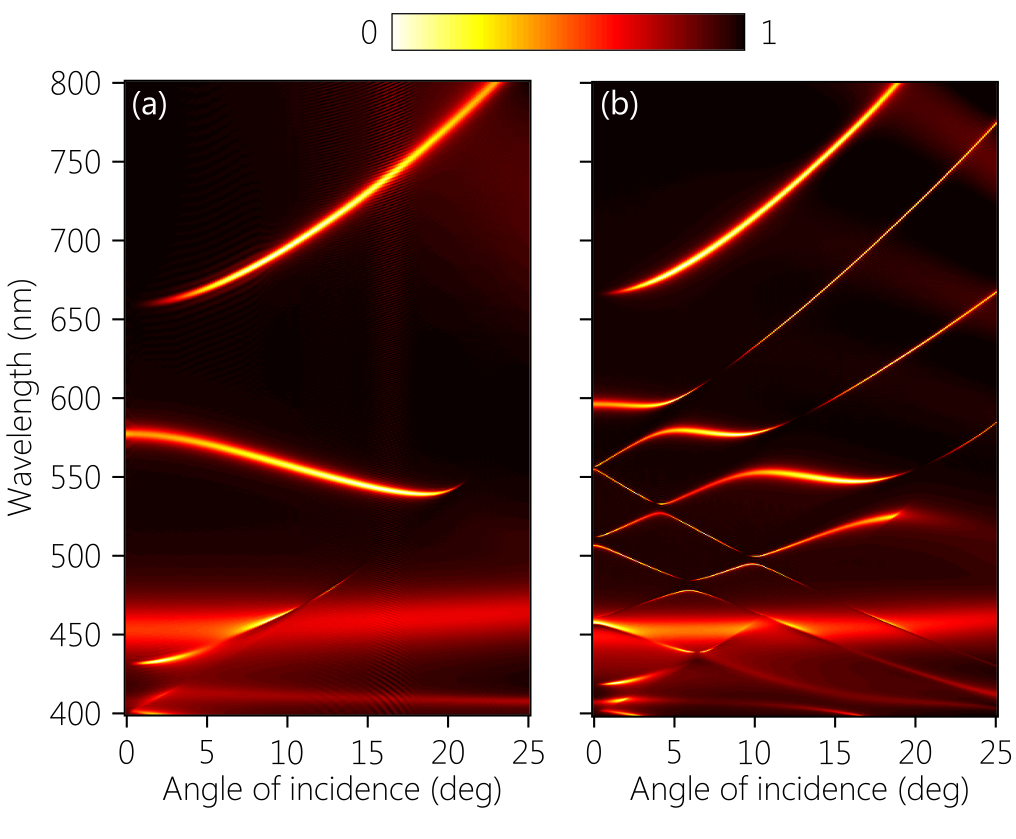}
    \caption{Band diagrams of the plasmonic grating (same as in Fig. 1.) coupled to a $\text{SiO}_2$ waveguiding layer with thicknesses of (a) 200 nm and (b) 800 nm. Optically thicker waveguide supports more guided modes allowing more interactions of resonances. } 
\label{fig:fig4} 
\end{figure}

As the linewidth of the resonance disappears, its Q factor diverges which manifests the formation of a BIC. 
Generally, the Q factor of a plasmonic mode is governed by the intrinsic metallic loss and the optical radiative loss. At (near) the BIC points, the radiation into free space is inhibited (minimized); therefore, the Q factors of the hybrid plasmonic-photonic modes around BICs are only metal loss-limited. The Q factors of the resonances around the three BICs discussed earlier are calculated by fitting the reflection coefficients to the Fano formula \cite{kivshar2017nature_review} and shown in Fig.~\ref{fig:fig3}. The Q factors of the symmetry-protected plasmonic and photonic BICs are depicted in Fig.~\ref{fig:fig3}(a) and \ref{fig:fig3}(b), respectively.
The nature of the mode is reflected in the electric fields plotted in the insets at an angle of $5 ^\circ$ and 697.1 nm wavelength for the plasmonic mode, Fig.~\ref{fig:fig3}(a), and $2 ^\circ$ and 540.1 nm for the photonic one in Fig.~\ref{fig:fig3}(b). The Q factor of the hybrid (Friedrich-Wintgen) BIC is shown in Fig.~\ref{fig:fig3}(c) with the type of interacting modes exchanged at the BIC point. The BIC appears at $11.4 ^\circ$,  and the hybridization is evident from the field plots given at the insets of Fig.~\ref{fig:fig3}(c) showing a plasmonic nature (at $5 ^\circ$ and 596.3 nm) and photonic nature (at $5 ^\circ$ and 630.3 nm).
At the BIC points, the radiative Q-factor diverges which causes the total Q to be limited by the loss in the system. Therefore, the total Q of the symmetry-protected  photonic BIC will diverge to infinity due to absence of loss, Fig.~\ref{fig:fig3}(c). However, both symmetry-protected  plasmonic BIC and Friedrich-Wintgen BIC will have finite, yet very high, Q values.

Varying the thickness of the waveguiding layer changes the number of the supported waveguide modes and strongly modifies the reflectance spectra. Fig.~\ref{fig:fig4} shows the reflection spectra vs. angle of incidence for the hybrid  structure with a top dielectric layer thickness of 200-nm [Fig.~\ref{fig:fig4}(a)] and 800-nm [Fig.~\ref{fig:fig4}(b)]. Several intriguing features can be noted from Fig.~\ref{fig:fig4}. First, the hybridization of the plasmonic and photonic modes allows for bandgap engineering depending on the number of modes supported by the system and their spectral overlap. From the modal analysis, a 200-nm thick silica slab waveguide surrounded by air would support a single TM mode at 600 nm wavelength. That explains the single band is appearing at Fig.~\ref{fig:fig4}(a). However, the 800-nm slab supports 3 TM modes at 600 nm which is manifested by the multiple interactions and avoided crossings depicted in Fig.~\ref{fig:fig4}(b) showing a more complicated behavior with multiple Friedrich-Wintgen BICs appearing at different $\theta$  values.
These results demonstrate the possibility of band gap engineering by designing the dispersion of either the photonic or the plasmonic component of the system. Moreover, the system supports slow light, not only at the band edge (around $\Gamma$-point)but also at multiple other $\theta$ values. 
More interestingly, the spectral bandwidth along which the slow light can be supported is substantially increased. We can clearly observe three distinct bands in Fig.~\ref{fig:fig4}(b) at which the slope of the resonance is almost zero indicating slow light. Unlike conventional plasmonic gratings where the slow light is supported at the band edge for only a single angle of incidence, here, the engineering of the band diagram with the mode hybridization allows extending this range significantly. The three slow light bands are centered at 550 nm, 577 nm, and 595 nm, with their angular bandwidths of $8 ^\circ$, $5 ^\circ$, and, $5 ^\circ$, respectively. Therefore, the proposed structure is capable of selecting the central angle at which the slow light exists, extending its angular span, and  providing means to select where the slow light is supported spectrally.

In conclusion, we have demonstrated the formation of BICs in systems with a realistic intrinsic loss. Due to the strong coupling and the destructive interference of the photonic and the plasmonic modes, the losses are significantly reduced due to suppression of radiation. Two different types of BICs are accessible in the hybrid plasmonic-photonic system due to the symmetry incompatibility with the radiation as well as the destructive reactions of resonances. A distinct characteristic of the Friedrich-Wintgen BIC achieved in this study is that it originates from the interference of two resonances with different natures. The system studied exhibits an exquisite set of phenomena including the formation of localized states with, in principle, infinite lifetimes, strong coupling with large Rabi splitting, optical bandgap engineering and slow light with broad spectral robustness. We believe that these hybrid structures are perfect candidates for low-threshold lasers, sharp spectral filters, enhancement of nonlinear phenomena as well as sensors.

\begin{acknowledgments}
The authors acknowledge the financial support by the U.S. Department of Energy (DOE), Office of Basic Energy Sciences (BES), Division of Materials Sciences and Engineering under Award DE-SC0017717.
\end{acknowledgments}

\bibliography{refs}
\end{document}